\documentclass[nofootinbib]{revtex4}
\usepackage{graphicx}
\usepackage{latexsym}
\usepackage{amsthm,amsmath,amssymb}
\usepackage{mathrsfs}
\def\be{\begin{equation}}
\def\ee{\end{equation}}
\def\bea{\begin{eqnarray}}
\def\eea{\end{eqnarray}}

\begin{document}

\hfill  USTC-ICTS/PCFT-25-07

\title{Spacetime perturbations and quasi-teleparallel gravity}

\author{Jian Gao}
\email{jiangao1108@mail.ustc.edu.cn}
\affiliation{Department of Astronomy, University of Science and Technology of China, Hefei, Anhui 230026, China}
\author{Yuxuan Kang}
\email{yxkang@mail.ustc.edu.cn}
\author{Mingzhe Li}
\email{limz@ustc.edu.cn}
\author{Yeheng Tong}
\email{yhtong19@mail.ustc.edu.cn}
\affiliation{Interdisciplinary Center for Theoretical Study, University of Science and Technology of China, Hefei, Anhui 230026, China}
\affiliation{Peng Huanwu Center for Fundamental Theory, Hefei, Anhui 230026, China}

\begin{abstract}

Gravity is identical to curved spacetime. It is manifested by the curvature of a Riemannian spacetime in general relativity but by torsion or non-metricity in teleparallel gravity models. In this paper, we apply these multiple options to the spacetime perturbation theory and seek the possibilities of representing the gravitation of the background and that of the perturbation in separate ways. We show that the perturbation around a Riemannian background can be described by torsion or non-metricity, so that we have teleparallel like actions for the perturbation. 

\end{abstract}

\maketitle

\section{introduction}

General relativity (GR) provided us a picture that gravitation is identical to curved spacetime, and is formulated by Riemannian geometry. With this, gravity is manifested by the curvature which is constructed from the Levi-Civita connection (or Christoffel symbol) and in turn from the metric. However, Riemannian geometry is not the unique approach to gravity theories. Sample theories based on non-Riemannian geometries include the Einstein-Cartan theory, the metric-affine theory, the teleparallel gravity, and so on. Usually in these theories, the connection is not limited to be the Levi-Civita type, and has no a priori dependence on the metric. For the teleparallel gravity \cite{Tele,Bahamonde:2021gfp,Krssak:2018ywd}, the curvature obtained from the connection vanishes and gravity is manifested by other geometric quantities: torsion in the metric teleparallel gravity (MTG), or non-metricity in the symmetric teleparallel gravity (STG) \cite{Nester:1998mp}. Within both frameworks one can build models equivalent to GR. This means treating the same thing in different pictures. 

Since GR is a highly non-linear theory, it is not easy to get exact solutions. In many cases, we have to resort to perturbation theory \cite{Mukhanov:1990me}, where physical quantities are divided into the background parts and their perturbations. Both the physical and background spacetimes are Riemannian geometric and solved the Einstein equation. More often than not, the background spacetime has high degree of symmetry and its form is assumed to be already known, for instances, the Schwarzschild solution and the Friedmann-Robertson-Walker universe. For the physical spacetime, however,  its exact form is not available. But we know it is close to the background and the deviation is treated as small perturbation. 

After separation, both the (curved) background spacetime and the perturbation around it manifest gravitational interactions. As mentioned above, gravity may be depicted in multiple ways: curvature, torsion and non-metricity. In this paper, we seek the possibilities of mixed patterns. More concretely, we ask the question whether the gravity identified with the background spacetime is represented by the curvature of Riemannian geometry, meanwhile the gravitation from the perturbation is represented by torsion or non-metricity, even though the whole physical spacetime (background plus perturbation) is fully Riemannian geometric? We will show that such pictures of quasi-teleparallel gravity for spacetime perturbations are possible. 

This paper is organized as follows: First we will briefly introduce the teleparallel gravity in Section II, then discuss the ways of formulating the spacetime perturbations with non-metricity and torsion respectively in Section III and IV, and conclude in Section V. 

\section{Teleparallel gravity}

Teleparallel gravity can be considered as constrained metric-affine models, so we start from the general metric-affine gravity theory \cite{Hehl:1994ue} where the metric tensor $g_{\mu\nu}$ and the affine connection $\hat{\Gamma}^{\rho}_{~\mu\nu}$ are considered to be independent. From the metric tensor one can construct the Levi-Civita connection,
\be\label{christoffel}
\Gamma^{\rho}_{~\mu\nu}=\frac{1}{2}g^{\rho\sigma}(\partial_{\mu}g_{\sigma\nu}+\partial_{\nu}g_{\mu\sigma}-\partial_{\sigma}g_{\mu\nu})~,
\ee
which is torsion free: $\Gamma^{\rho}_{~\mu\nu}=\Gamma^{\rho}_{~\nu\mu}$, and metric-compatible: $\nabla_{\rho}g_{\mu\nu}=\partial_{\rho}g_{\mu\nu}-\Gamma^{\alpha}_{~\rho\mu}g_{\alpha\nu}-\Gamma^{\alpha}_{~\rho\nu}g_{\mu\alpha}=0$. Usually this is not the case for the general affine connection $\hat{\Gamma}^{\rho}_{~\mu\nu}$, its lack of these properties are characterized by the torsion tensor $T^{\rho}_{~\mu\nu}\equiv  \hat{\Gamma}^{\rho}_{~\mu\nu}-\hat{\Gamma}^{\rho}_{~\nu\mu}$ and the non-metricity tensor $Q_{\rho\mu\nu}\equiv \hat{\nabla}_{\rho}g_{\mu\nu}=\partial_{\rho}g_{\mu\nu}-\hat{\Gamma}^{\alpha}_{~\rho\mu}g_{\alpha\nu}-\hat{\Gamma}^{\alpha}_{~\rho\nu}g_{\mu\alpha}$. With these definitions, the distortion tensor $C^{\rho}_{~\mu\nu}\equiv \hat{\Gamma}^{\rho}_{~\mu\nu}-\Gamma^{\rho}_{~\mu\nu}$, which measures the difference between $\hat{\Gamma}^{\rho}_{~\mu\nu}$ and $\Gamma^{\rho}_{~\mu\nu}$, can be decomposed as 
\be\label{distortion}
C^{\rho}_{~\mu\nu}= K^{\rho}_{~\mu\nu} +L^{\rho}_{~\mu\nu} ~,
\ee
where
\be\label{ktensor}
K^{\rho}_{~\mu\nu}= \frac{1}{2}(T^{\rho}_{~\mu\nu}-T_{\mu\nu}^{~~\rho}-T_{\nu\mu}^{~~\rho})~,
\ee
is the contortion tensor, and 
\be\label{ltensor}
L^{\rho}_{~\mu\nu}=\frac{1}{2}(Q^{\rho}_{~\mu\nu}-Q_{\mu\nu}^{~~\rho}-Q_{\nu\mu}^{~~\rho})~,
\ee
is the disformation tensor. To get these relations, we have used the metric $g_{\mu\nu}$ and its inverse to lower and rise the tensor indices, and considered the properties that  the torsion tensor is antisymmetric $T^{\rho}_{~\mu\nu}=-T^{\rho}_{~\nu\mu}$ but the non-metricity tensor is symmetric $Q_{\rho\mu\nu}=Q_{\rho\nu\mu}$ under the interchange of the last two indices. Besides the metric, the contortion tensor only depends on torsion, but the disformation tensor only depends on non-metricity. 
The Riemann curvature tensor can be obtained from the connection and its derivatives. Since there are two kinds of connections, we will have two different curvature tensors. One is from the general connection, 
 \be
 \hat{R}^{\rho}_{~\sigma\mu\nu}=\partial_{\mu}\hat{\Gamma}^{\rho}_{~\nu\sigma}-\partial_{\nu}\hat{\Gamma}^{\rho}_{~\mu\sigma}+\hat{\Gamma}^{\rho}_{~\mu\alpha}\hat{\Gamma}^{\alpha}_{~\nu\sigma}-\hat{\Gamma}^{\rho}_{~\nu\alpha}\hat{\Gamma}^{\alpha}_{~\mu\sigma}~,
 \ee
and another is from the Levi-Civita connection,
  \be
 R^{\rho}_{~\sigma\mu\nu}=\partial_{\mu}\Gamma^{\rho}_{~\nu\sigma}-\partial_{\nu}\Gamma^{\rho}_{~\mu\sigma}+\Gamma^{\rho}_{~\mu\alpha}\Gamma^{\alpha}_{~\nu\sigma}-\Gamma^{\rho}_{~\nu\alpha}\Gamma^{\alpha}_{~\mu\sigma}~.
 \ee
The difference between these two curvature tensors depends on the distortion tensor $C^{\rho}_{~\mu\nu}$ in Eq. (\ref{distortion}) and in turn on the torsion and non-metricity, 
  \be\label{riemanndifference}
 R^{\rho}_{~\sigma\mu\nu}=\hat{R}^{\rho}_{~\sigma\mu\nu}-\nabla_{\mu}C^{\rho}_{~\nu\sigma}+\nabla_{\nu}C^{\rho}_{~\mu\sigma}-C^{\rho}_{~\mu\alpha}C^{\alpha}_{~\nu\sigma}+C^{\rho}_{~\nu\alpha}C^{\alpha}_{~\mu\sigma}~,
 \ee
here again the covariant derivative operator $\nabla_{\mu}$ is associated with the Levi-Civita connection. 
Then we have the Ricci tensor 
\be
R_{\mu\nu}=\hat{R}_{\mu\nu}-\nabla_{\rho}C^{\rho}_{~\nu\mu}+\nabla_{\nu}C^{\rho}_{~\rho\mu}-C^{\rho}_{~\rho\alpha}C^{\alpha}_{~\nu\mu}+C^{\rho}_{~\nu\alpha}C^{\alpha}_{~\rho\mu}~,
 \ee
and the curvature scalar 
\be\label{curvaturescalar}
R\equiv g^{\mu\nu}R_{\mu\nu}=g^{\mu\nu}\hat{R}_{\mu\nu}+\nabla_{\mu}(C_{\rho}^{~\rho\mu}-C^{\mu\rho}_{~~~\rho})-C^{\rho}_{~\rho\alpha}C^{\alpha\mu}_{~~~\mu}+C^{\rho\mu}_{~~~\alpha}C^{\alpha}_{~\rho\mu}~.
\ee

Teleparallel gravity can be obtained by imposing on the general metric-affine models the teleparallelism constraint, i.e., the curvature tensor from the general affine connection vanishes: $\hat{R}^{\rho}_{~\sigma\mu\nu}=0$. So that the curvature tensor from the Levi-Civita connection is totally determined by the distortion tensor and the metric itself,  
\be
R^{\rho}_{~\sigma\mu\nu}=-\nabla_{\mu}C^{\rho}_{~\nu\sigma}+\nabla_{\nu}C^{\rho}_{~\mu\sigma}- C^{\rho}_{~\mu\alpha}C^{\alpha}_{~\nu\sigma}+C^{\rho}_{~\nu\alpha}C^{\alpha}_{~\mu\sigma}~.
\ee
The curvature scalar, which now takes the form 
\be\label{tele}
R=\nabla_{\mu}(C_{\rho}^{~\rho\mu}-C^{\mu\rho}_{~~~\rho})-C^{\rho}_{~\rho\alpha}C^{\alpha\mu}_{~~~\mu}+C^{\rho\mu}_{~~~\alpha}C^{\alpha}_{~\rho\mu}~,
\ee
plays an important role in general relativity where the Einstein-Hilbert action for gravity is given by
\be
S_{\rm GR}=\frac{1}{2}\int d^4 x\sqrt{-g} R~,
\ee
here $g$ is the determinant of the metric $g_{\mu\nu}$ and the unit $8\pi G=1$ was adopted. 

As mentioned before, there are two kinds of frequently studied teleparallel gravity models: MTG and STG. In the MTG model, besides the teleparallelism constraint, the metric-compatibility is also required for $\hat{\Gamma}^{\rho}_{~\mu\nu}$, i.e., $Q_{\rho\mu\nu}=0$, so that $C^{\rho}_{~\mu\nu}=K^{\rho}_{~\mu\nu}$. In this case, the curvature scalar from the Levi-Civita connection in Eq. (\ref{tele}) becomes
\be\label{tele1}
R=T_{\mu}T^{\mu}-\frac{1}{4}T_{\rho\sigma\mu}T^{\rho\sigma\mu}-\frac{1}{2}T_{\rho\sigma\mu}T^{\sigma\rho\mu}+2\nabla_{\mu}T^{\mu}\equiv \mathbb{T}+2\nabla_{\mu}T^{\mu}~,
\ee
where $T_{\mu}=T^{\nu}_{~\nu\mu}$ is the torsion vector and the defined $\mathbb{T}$ is a torsion scalar and is in quadratic form of the torsion tensor. With these, one can construct a model equivalent to general relativity within the framework of MTG. Such a model depends heavily on torsion tensor (denoted by $T$) and is equivalent to general relativity, we may call it TGR model. It has the action
\be\label{tgr}
S_{\rm TGR}=\frac{1}{2}\int d^4x\sqrt{-g} \mathbb {T}~,
\ee
it is the same as $S_{\rm GR}$ after integrating out the total derivative term in Eq. (\ref{tele1}). Other MTG models are considered as extensions to TGR, such as the $f(\mathbb{T})$ model \cite{Yang:2010ji, Cai:2015emx, Krssak:2015oua, Bahamonde:2022ohm} and the Nieh-Yan modified Teleparallel Gravity (NYTG) model \cite{Li:2020xjt, Li:2021wij, Li:2023fto,Zhang:2024vfw}, and so on. The MTG model gives a picture that gravity is manifested by torsion in stead of curvature. Please note that in MTG, the affine connection $\hat{\Gamma}^{\rho}_{~\mu\nu}$ or the torsion tensor are not fundamental variables. The teleparallelism constraint determines that $\hat{\Gamma}^{\rho}_{~\mu\nu}$ can neither have a general form nor be varied freely when using the variation principle. Furthermore, torsion is required to be existent, so that one cannot simply set all the components of $\hat{\Gamma}^{\rho}_{~\mu\nu}$ to zero to fit the teleparallelism constraint. To find the true building blocks of the MTG models, it is convenient to make use of the tetrad formulation. With this language the metric is built from the the tetrad (or veilbein) $e^{a}_{\mu}$ through the relation $g_{\mu\nu}=e^{a}_{\mu}e^{b}_{\nu}\eta_{ab}$, here $\eta_{ab}={\rm diag}(-1, +1,+1, +1)$ is the Minkowski metric and the Latin letters $a, b, ...$ are the Lorentz indices, used to denote the tensor components at the local flat space.  The affine connection $\hat{\Gamma}^{\rho}_{~\mu\nu}$ is constructed by the tetrad and the spin connection $\hat{\omega}^a_{~b\mu}$ as $\hat{\Gamma}^{\rho}_{~\mu\nu}=\theta^{\rho}_{a}(\partial_{\mu}e^a_{\nu}+\hat{\omega}^a_{~b\mu}e^b_{\nu})$, here  $\theta^{\rho}_{a}$ is the inverse of the tetrad: $\theta^{\rho}_{a}e^a_{\sigma}=\delta^{\rho}_{\sigma}$ and $\theta^{\rho}_{a}e^b_{\rho}=\delta^{b}_{a}$. Then with the definition $T^{\rho}_{~\mu\nu}=  \hat{\Gamma}^{\rho}_{~\mu\nu}-\hat{\Gamma}^{\rho}_{~\nu\mu}$, it is straightforwardly to obtain the expression of the torsion tensor
\be
T^{\rho}_{~\mu\nu}=\theta^{\rho}_{a}(\partial_{\mu}e^a_{\nu}-\partial_{\nu}e^a_{\mu}+\hat{\omega}^a_{~b\mu}e^b_{\nu}-\hat{\omega}^a_{~b\nu}e^b_{\mu})~.
\ee
The spin connection, under the teleparallelism constraint and the requirement of metric-compatibility, can be expressed as $\hat{\omega}^{a}_{~b\mu}=(\Lambda^{-1})^a_{~c}\partial_{\mu}\Lambda^c_{~b}$, here $\Lambda^a_{~b}$ is the position dependent Lorentz transform. For the TGR model (\ref{tgr}), it is safely to adopt the 
Weitzenb\"{o}ck condition $\hat{\omega}^a_{~b\mu}=0$, so that $T^{\rho}_{~\mu\nu}=\theta^{\rho}_{a}(\partial_{\mu}e^a_{\nu}-\partial_{\nu}e^a_{\mu})$, and the TGR model itself can be considered as a pure tetrad model, where only the tetrads are considered as fundamental fields and the torsion which contains derivatives of tetrad is considered as the strength field. 

In the STG model, the general affine connection $\hat{\Gamma}^{\rho}_{~\mu\nu}$ is constrained to be torsionless besides the teleparallelism constraint, so that $C^{\rho}_{~\mu\nu}=L^{\rho}_{~\mu\nu}$. In this case, the curvature scalar from the Levi-Civita connection in Eq. (\ref{tele}) becomes
\be\label{tele2}
R=-\frac{1}{4}Q_{\rho\sigma\mu}Q^{\rho\sigma\mu}+\frac{1}{2}Q_{\rho\sigma\mu}Q^{\sigma\mu\rho}+\frac{1}{4}Q_{\mu}Q^{\mu}-\frac{1}{2}Q_{\mu}\bar{Q}^{\mu}+\nabla_{\mu}(\bar{Q}^{\mu}-Q^{\mu})\equiv \mathbb{Q}+\nabla_{\mu}(\bar{Q}^{\mu}-Q^{\mu})~,
\ee
where $Q^{\mu}=Q^{\mu\nu}_{~~~\nu}~,~\bar{Q}^{\mu}=Q^{\nu~\mu}_{~\nu}$ are two non-metricity vectors, and the defined $\mathbb{Q}$ is a non-metricity scalar which is in quadratic form of the non-metricity tensor. Similarly, one can construct a model equivalent to general relativity within the framework of STG. Such a model depends heavily on non-metricity tensor (denoted by $Q$) and is equivalent to general relativity, we may call it QGR model. It has the action
\be\label{qgr}
S_{\rm QGR}=\frac{1}{2}\int d^4x\sqrt{-g} \mathbb {Q}~.
\ee
It is the same as $S_{\rm GR}$ after integrating out the total derivative terms in Eq. (\ref{tele2}). Other symmetric teleparallel gravity models are considered as extensions to QGR, for example the $f(\mathbb{Q})$ model \cite{Zhao:2021zab, Zhao:2024kri, Lu:2019hra, Hohmann:2021ast,Frusciante:2021sio, Rao:2024ncj} and the model with parity-violating extensions \cite{Li:2022vtn,Zhang:2023scq}. The STG model provides a picture that gravity is ascribed to the non-metricity. In this picture, the metric is fundamental. For the QGR model (\ref{qgr}), it is safely to adopt the so called coincident gauge where $\hat{\Gamma}^{\rho}_{~\mu\nu}=0$, so that $Q_{\rho\mu\nu}=\partial_{\rho}g_{\mu\nu}$ and the model is a pure metric one.  

\section{spacetime perturbation via non-metricity}

From now on, we address to the question of how to formulate gravitations from the background and the perturbation with separate pictures in the perturbation theory. We first consider this problem within the metric formulation. With this language, we have a metric $g_{\mu\nu}$ for the physical spacetime and its Levi-Civita connection $\Gamma^{\rho}_{~\mu\nu}$ defined in Eq. (\ref{christoffel}).  At the same time we have a metric $\bar{g}_{\mu\nu}$ for the background spacetime and its corresponding Levi-Civita connection $\bar{\Gamma}^{\rho}_{~\mu\nu}=\frac{1}{2}\bar{g}^{\rho\sigma}(\partial_{\mu}\bar{g}_{\sigma\nu}+\partial_{\nu}\bar{g}_{\mu\sigma}-\partial_{\sigma}\bar{g}_{\mu\nu})$.  
Both the physical and background spacetimes are of Riemannian geometries, and manifest respective gravitational interactions through respective curvatures, $R(\Gamma)$ and $\bar{R}(\bar{\Gamma})$. The spacetime perturbation arise from the difference between these two metrics, this leads to the mismatch between $g_{\mu\nu}$ and $\bar{\Gamma}^{\rho}_{~\mu\nu}$, or between $\bar{g}_{\mu\nu}$ and $\Gamma^{\rho}_{~\mu\nu}$. 

From the viewpoint of metric-affine theory, it is convenient to choose the hatted connection in previous section as the background Levi-Civita connection: $\hat{\Gamma}^{\rho}_{~\mu\nu}=\bar{\Gamma}^{\rho}_{~\mu\nu}$, so that the torsion vanishes but the non-metricity $Q_{\rho\mu\nu}=\partial_{\rho}g_{\mu\nu}-\bar{\Gamma}^{\alpha}_{~\rho\mu}g_{\alpha\nu}-\bar{\Gamma}^{\alpha}_{~\rho\nu}g_{\mu\alpha}$ does not. The distortion tensor $C^{\rho}_{~\mu\nu}= \bar{\Gamma}^{\rho}_{~\mu\nu}-\Gamma^{\rho}_{~\mu\nu}=L^{\rho}_{~\mu\nu}$ is determined by the non-metricity tensor as indicated in Eq. (\ref{ltensor}). 
Different from the full STG model, here the curvature tensor $\bar{R}^{\rho}_{~\sigma\mu\nu}$ induced by $\bar{\Gamma}^{\rho}_{~\mu\nu}$ represents the curvature of the background spacetime and does not vanish unless the background is flat. Its exact form is assumed to be already known in the perturbation theory. 
Then we have the Riemann curvature tensor for the physical spacetime
 \be\label{curvaturesepa}
 R^{\rho}_{~\sigma\mu\nu}=\bar{R}^{\rho}_{~\sigma\mu\nu}-\nabla_{\mu}L^{\rho}_{~\nu\sigma}+\nabla_{\nu}L^{\rho}_{~\mu\sigma}-L^{\rho}_{~\mu\alpha}L^{\alpha}_{~\nu\sigma}+L^{\rho}_{~\nu\alpha}L^{\alpha}_{~\mu\sigma}\equiv \bar{R}^{\rho}_{~\sigma\mu\nu}+\delta R^{\rho}_{~\sigma\mu\nu}~,
 \ee
and the Ricci tensor 
\be
R_{\mu\nu}=\bar{R}_{\mu\nu}-\nabla_{\rho}L^{\rho}_{~\nu\mu}+\nabla_{\nu}L^{\rho}_{~\rho\mu}-L^{\rho}_{~\rho\alpha}L^{\alpha}_{~\nu\mu}+L^{\rho}_{~\nu\alpha}L^{\alpha}_{~\rho\mu}\equiv \bar{R}_{\mu\nu}+\delta R_{\mu\nu}~.
 \ee
These have been written in the form of separating background and perturbation. The perturbations $\delta R^{\rho}_{~\sigma\mu\nu}$ and $\delta R_{\mu\nu}$ are ascribed to non-metricity. 

The curvature scalar for the physical spacetime changes to 
\be
R=g^{\mu\nu}\bar{R}_{\mu\nu}+\nabla^{\rho}L^{\mu}_{~\mu\rho}-\nabla_{\rho}L^{\rho\mu}_{~~~\mu}+L^{\rho\mu}_{~~~\nu}L^{\nu}_{~\rho\mu}-L^{\rho}_{~\rho\nu}L^{\nu\mu}_{~~~\mu}=g^{\mu\nu}\bar{R}_{\mu\nu}+ \mathbb{Q}+\nabla_{\mu}(\bar{Q}^{\mu}-Q^{\mu})~, 
\ee
where the non-metricity scalar $\mathbb{Q}$ is exactly the same as that in the STG model defined in Eq. (\ref{tele2}). 
If the underlying theory for the physical spacetime is GR, the Einstein-Hilbert action of gravity $S=(1/2)\int d^4x\sqrt{-g}R$ is rewritten as 
\be\label{QGRlike}
S=\frac{1}{2}\int d^4x\sqrt{-g}(g^{\mu\nu}\bar{R}_{\mu\nu}+ \mathbb{Q})~,
\ee
after removing the total derivative terms. Now we obtained a QGR like action for the spacetime perturbations. If the background spacetime is flat, $\bar{g}_{\mu\nu}=\eta_{\mu\nu}$, $\bar{\Gamma}^{\rho}_{~\mu\nu}=0$, $\bar{R}_{\mu\nu}=0$ and $Q_{\rho\mu\nu}=\partial_{\rho}g_{\mu\nu}$, the full action becomes $S=(1/2)\int d^4x \sqrt{-g}\mathbb{Q}$, going back to the action of the QGR model (\ref{qgr}) under the coincident gauge. 

Please note that in the action integral (\ref{QGRlike}), spacetime perturbation is not totally described by the second term $\sqrt{-g}\mathbb{Q}$, the first term also contributes to the action for perturbation because the perturbation arises from the deviation of $g_{\mu\nu}$ from the background metric $\bar{g}_{\mu\nu}$ and the product $\sqrt{-g}g^{\mu\nu}$ itself contains perturbation. Now we consider the expansion of the action as the perturbative series. Firstly, we use the exponential map to describe the deviation of $g_{\mu\nu}$ from the background metric:
\be\label{map}
g_{\mu\nu}=\left(e^{\epsilon}\right)^{\rho}_{~\mu}\left(e^{\epsilon}\right)^{\sigma}_{~\nu}\bar{g}_{\rho\sigma}~,
\ee
where $\epsilon$ is a small matrix and its elements $\epsilon^{\mu}_{~\nu}$ are considered as the basic perturbation variables. This parameterization of the perturbation is not conventional, but for our purposes in this paper it has some advantages relative to the conventional parameterization of the metric perturbation, such as $g_{\mu\nu}=\bar g_{\mu\nu}+h_{\mu\nu}$ or the exponential form $g_{\mu\nu}= \bar g_{\mu\rho}\left(e^h\right)^{\rho}_{~\nu}$. With the parameterization of Eq. (\ref{map}), $e^{\epsilon}$ has the meaning of transfer matrix for the  map between the physical spacetime and the background. It is the same with the parameterization of the tetrad perturbation, which will be discussed in the next section. In addition, this parameterization automatically guaranteed the symmetric property of $g_{\mu\nu}$ as long as the background metric $\bar g_{\rho\sigma}$ is symmetric. As a comparison, in the conventional method, one should presuppose the symmetry of $h_{\mu\nu}$. But with the parameterization of Eq. (\ref{map}), the linear perturbation to the metric is $\epsilon_{\mu\nu}+\epsilon_{\nu\mu}$ with $\epsilon_{\mu\nu}\equiv \bar{g}_{\mu\rho}\epsilon^{\rho}_{~\nu}$, which corresponds to $h_{\mu\nu}$ in the conventional parameterization.
With Eq. (\ref{map}), one can obtain that
\be
\sqrt{-g}g^{\mu\nu}=\sqrt{-\bar{g}}e^{\rm Tr\epsilon} \left(e^{-\epsilon}\right)^{\mu}_{~\rho}\left(e^{-\epsilon}\right)^{\nu}_{~\sigma}\bar{g}^{\rho\sigma}~.
\ee
Secondly, we know $\mathbb{Q}$ is in quadratic form of the non-metricity tensor $Q_{\rho\mu\nu}$ and the latter is at least a first order perturbation quantity, so $\mathbb{Q}$ should be at least a second order quantity. With these we expand the action (\ref{QGRlike}) up to the second order: $S=S^{(0)}+S^{(1)}+S^{(2)}+\dots$, with
\be\label{action}
S^{(0)}=\frac{1}{2}\int d^4 x\sqrt{-\bar{g}}\bar{R}~,~S^{(1)}=-\int d^4 x\sqrt{-\bar{g}}\bar{G}^{\mu}_{~\nu} \epsilon^{\nu}_{~\mu}~,
\ee
and 
\be
S^{(2)}=\frac{1}{2}\int d^4 x\sqrt{-\bar{g}}[(\bar{R}_{\mu\nu}\bar{g}^{\rho\sigma}+\bar{R}^{\rho}_{~\nu}\delta^{\sigma}_{~\mu}-\bar{R}^{\sigma}_{~\nu}\delta^{\rho}_{~\mu}-\bar{G}^{\sigma}_{~\nu}\delta^{\rho}_{~\mu})\epsilon^{\mu}_{~\rho}\epsilon^{\nu}_{~\sigma}+\mathbb{Q}]~.
\ee
In above equations, the lowering, raising and contractions of the background spacetime indices are done by the background metric $\bar{g}_{\mu\nu}$ and its inverse, so $\bar{R}=\bar{g}^{\mu\nu}\bar{R}_{\mu\nu}~,~\bar{R}^{\nu}_{~\sigma}=\bar{g}^{\mu\nu}\bar{R}_{\mu\sigma}$, and $\bar{G}^{\mu}_{~\nu}=\bar{R}^{\mu}_{~\nu}-(1/2)\bar{R}\delta^{\mu}_{~\nu}$ is the background Einstein tensor. The zeroth order action $S^{(0)}$ leads to the Einstein field equation for the background: $\bar{G}^{\mu}_{~\nu}=\bar{T}^{\mu}_{~\nu}$ and $\bar{T}^{\mu}_{~\nu}$ is the energy-momentum tensor when matter couplings are included. At the same time, matter couplings contribute a linear order term $\int d^4 x\sqrt{-\bar{g}}\bar{T}^{\mu}_{~\nu} \epsilon^{\nu}_{~\mu}$, which cancels out $S^{(1)}$ in Eq. (\ref{action}) when the background equation holds. The same reason is valid for $S^{(2)}$ in which the term involves $\bar{G}^{\sigma}_{~\nu}$ would be canceled out by the matter action. So the quadratic action for the spacetime perturbation is 
\be\label{action2}
S^{(2)}=\frac{1}{2}\int d^4 x\sqrt{-\bar{g}}[(\bar{R}_{\mu\nu}\bar{g}^{\rho\sigma}+\bar{R}^{\rho}_{~\nu}\delta^{\sigma}_{~\mu}-\bar{R}^{\sigma}_{~\nu}\delta^{\rho}_{~\mu})\epsilon^{\mu}_{~\rho}\epsilon^{\nu}_{~\sigma}+\mathbb{Q}]~.
\ee
The quadratic action is a central element for the linear perturbation theory. One can see from Eq. (\ref{action2}) that all the derivatives of perturbations are contained in the non-metricity scalar $\mathbb{Q}$, The background curvature, appearing as coefficients, merely contributes to the ``potential" of $\epsilon^{\mu}_{~\nu}$. The dynamics of the perturbation is mainly governed by the non-metricity.

 \section{spacetime perturbation via torsion}

Now we turn to the question of depicting the perturbation with torsion. This is not easy in terms of the metric formulation used in the previous section, because both the affine connections $\Gamma^{\rho}_{~\mu\nu}$ and $\bar{\Gamma}^{\rho}_{~\mu\nu}$ are Christoffel symbols and torsion free. So we switch to the tetrad formulation. 

We use the tetrad $e^{a}_{\mu}$ (and its inverse $\theta^{\mu}_a$) to denote the ``square root" of the physical spacetime metric, i.e., $g_{\mu\nu}=e^{a}_{\mu}e^{b}_{\nu}\eta_{ab}$. It matches the spin connection $\omega^a_{~b\mu}$. That is, with respect to 
$e^{a}_{\mu}$, the spin connection $\omega^a_{~b\mu}$ is torsion free:
\be\label{torsionfree}
\partial_{\mu}e^a_{\nu}-\partial_{\nu}e^a_{\mu}+\omega^a_{~b\mu}e^b_{\nu}-\omega^a_{~b\nu}e^b_{\mu}=0~, 
\ee
and metric-compatible: $\omega_{ab\mu}=-\omega_{ba\mu}$, here the Lorentz indices are lowered by the Minkowski metric. 
The affine connection $\Gamma^{\rho}_{~\mu\nu}=\theta^{\rho}_{a}(\partial_{\mu}e^a_{\nu}+\omega^a_{~b\mu}e^b_{\nu})$ is just the Levi-Civita connection of the metric $g_{\mu\nu}$.

At the same time, we have the background tetrad $\bar{e}^{a}_{\mu}$ and its matched spin connection $\bar{\omega}^a_{~b\mu}$. Their exact forms are assumed to be known. 
Correspondingly, we have the background metric $\bar{g}_{\mu\nu}=\bar{e}^{a}_{\mu}\bar{e}^{b}_{\nu}\eta_{ab}$ 
and its Levi-Civita connection $\bar{\Gamma}^{\rho}_{~\mu\nu}=\bar{\theta}^{\rho}_{a}(\partial_{\mu}\bar{e}^a_{\nu}+\bar{\omega}^a_{~b\mu}\bar{e}^b_{\nu})$. 
However, the physical spacetime tetrad $e^{a}_{\mu}$ does not match the background spin connection $\bar{\omega}^a_{~b\mu}$\footnote{Of course, this mismatching also happens between $\bar{e}^{a}_{\mu}$ and $\omega^a_{~b\mu}$}. This mismatching gives rise to the torsion:
\be\label{torsion}
T^a_{~\mu\nu}=\partial_{\mu}e^a_{\nu}-\partial_{\nu}e^a_{\mu}+\bar{\omega}^a_{~b\mu}e^b_{\nu}-\bar{\omega}^a_{~b\nu}e^b_{\mu}~, 
\ee
and relates to the torsion tensor mentioned at Section II via the relation: $T^{\rho}_{~\mu\nu}=\theta^{\rho}_aT^a_{~\mu\nu}$. 
Please note that the affine connection from the mismatched pair $(e^a_{\mu}~,~\bar{\omega}^a_{~b\mu})$ is $\tilde{\Gamma}^{\rho}_{~\mu\nu}=\theta^{\rho}_{a}(\partial_{\mu}e^a_{\nu}+\bar{\omega}^a_{~b\mu}e^b_{\nu})$, which is neither the Levi-Civita affine connection $\Gamma^{\rho}_{~\mu\nu}$ for the physical spacetime metric nor the Levi-Civita affine connection $\bar{\Gamma}^{\rho}_{~\mu\nu}$ for the background metric. But it is precisely the antisymmetry of this affine connection that gives rise to the non-vanishing torsion: $T^{\rho}_{~\mu\nu}=  \tilde{\Gamma}^{\rho}_{~\mu\nu}-\tilde{\Gamma}^{\rho}_{~\nu\mu}$. 

Now we introduce $M^a_{~b\mu}$ to measure the difference between the spin connections $\bar{\omega}^a_{~b\mu}$ and $\omega^a_{~b\mu}$, 
\be\label{mmunu}
M^a_{~b\mu}\equiv \bar{\omega}^a_{~b\mu}-\omega^a_{~b\mu}~.
\ee
Since both $\bar{\omega}_{ab\mu}$ and $\omega_{ab\mu}$ are antisymmetric under the interchange of the first two Lorentz indices, so is the $M$-tensor: $M_{ab\mu}=-M_{ba\mu}$. 

From the torsion defined in Eq. (\ref{torsion}) and the torsion free equation (\ref{torsionfree}) for $\omega^{a}_{~b\mu}$, one can obtain that the torsion is determined only by the $M$-tensor,
\be
T^a_{~\mu\nu}=M^a_{~b\mu}e^b_{\nu}-M^a_{~b\nu}e^b_{\mu}=M^a_{~\nu\mu}-M^a_{~\mu\nu}~,~T^{\rho}_{~\mu\nu}=M^{\rho}_{~\nu\mu}-M^{\rho}_{~\mu\nu}~.
\ee
In the second equation above we have defined the tensors $M^{\rho}_{~\sigma\mu}=\theta^{\rho}_a e^b_{\sigma}M^a_{~b\mu}$ and so on.
Combining it with the antisymmetric property, $M_{\rho\sigma\mu}=-M_{\sigma\rho\mu}$, it is not difficult to express the $M$-tensor with torsion as 
\be\label{mtensor1}
M^{\rho}_{~\mu\nu}=-\frac{1}{2}(T^{\rho}_{~\mu\nu}+T_{\mu\nu}^{~~\rho}+T_{\nu\mu}^{~~\rho})~.
\ee

The curvature for the background is determined by the spin connection $\bar{\omega}^a_{~b\mu}$ alone, 
\be
\bar{R}^a_{~b\mu\nu}=\partial_{\mu}\bar{\omega}^a_{~b\nu}-\partial_{\nu}\bar{\omega}^a_{~b\mu}+\bar{\omega}^a_{~c\mu}\bar{\omega}^c_{~b\nu}
-\bar{\omega}^a_{~c\nu}\bar{\omega}^c_{~b\mu}~, 
\ee
there is a same formula for the physical spacetime curvature $R^a_{~b\mu\nu}$ with $\omega^{a}_{~b\mu}$, i.e., $R^a_{~b\mu\nu}=\partial_{\mu}\omega^a_{~b\nu}-\partial_{\nu}\omega^a_{~b\mu}+\omega^a_{~c\mu}\omega^c_{~b\nu}-\omega^a_{~c\nu}\omega^c_{~b\mu}$. Now with $M^a_{~b\mu}$ defined in Eq. (\ref{mmunu}), one can obtain the following relation,
\be\label{curvatureform}
R^a_{~b\mu\nu}=\bar{R}^a_{~b\mu\nu}-D_{\mu}M^a_{~b\nu}+D_{\nu}M^a_{~b\mu}-M^a_{~c\mu}M^c_{~b\nu}+M^a_{~c\nu}M^c_{~b\mu}\equiv \bar{R}^a_{~b\mu\nu}+\delta R^a_{~b\mu\nu}~,
\ee
where the covariant derivative $D_{\mu}$ is associated with both the spin connection $\omega^a_{~b\mu}$ and the affine connection $\Gamma^{\rho}_{~\mu\nu}$, for example, 
$D_{\mu}M^a_{~b\nu}=\partial_{\mu}M^a_{~b\nu}+\omega^a_{~c\mu}M^c_{~b\nu}-\omega^c_{~b\mu}M^a_{~c\nu}-\Gamma^{\rho}_{~\mu\nu}M^a_{~b\rho}$. According to the ``tetrad postulate", the covariant derivative of the tetrad vanishes identically: $D_{\mu}e^a_{\nu}=\partial_{\mu}e^a_{\nu}+\omega^a_{~b\mu}e^b_{\nu}-\Gamma^{\rho}_{~\mu\nu}e^a_{\rho}=0$. This is consistent with the relation: $\Gamma^{\rho}_{~\mu\nu}=\theta^{\rho}_{a}(\partial_{\mu}e^a_{\nu}+\omega^a_{~b\mu}e^b_{\nu})$. It is easy to prove that the covariant derivative of the inverse of the tetrad also vanishes identically, $D_{\mu}\theta^{\nu}_a=0$. This fact facilitates us to freely move the tetrad and its inverse in and out of the covariant derivatives. The equation (\ref{curvatureform}) has been written in the form of separating background and perturbation. One can see that the perturbation to the curvature, $\delta\bar{R}^a_{~b\mu\nu}$, is ascribed to $M^a_{~b\mu}$ which comes from torsion according to Eq. (\ref{mtensor1}). 

The Riemann curvature tensor with all components are labeled by spacetime indices can be obtained through Eq. (\ref{curvatureform}) with the help of the tetrad and its inverse, 
\be 
R^{\rho}_{~\sigma\mu\nu}=\theta^{\rho}_a e^b_{\sigma}R^a_{~b\mu\nu}=\theta^{\rho}_a e^b_{\sigma}\bar{R}^a_{~b\mu\nu}-D_{\mu}M^{\rho}_{~\sigma\nu}+D_{\nu}M^{\rho}_{~\sigma\mu}
-M^{\rho}_{~\alpha\mu}M^{\alpha}_{~\sigma\nu}+M^{\rho}_{~\alpha\nu}M^{\alpha}_{~\sigma\mu}~.
\ee
We should note that $\theta^{\rho}_a e^b_{\sigma}\bar{R}^a_{~b\mu\nu}\neq \bar{R}^{\rho}_{~\sigma\mu\nu}$, the latter will be defined as $\bar{R}^{\rho}_{~\sigma\mu\nu}=\bar{\theta}^{\rho}_a \bar{e}^b_{\sigma}\bar{R}^a_{~b\mu\nu}$.  
Then we have the Ricci tensor after taking the trace of the Riemann curvature tensor:
\be
R_{\mu\nu}=\theta^{\rho}_a e^b_{\mu}\bar{R}^a_{~b\rho\nu}-D_{\rho}M^{\rho}_{~\mu\nu}+D_{\nu}M^{\rho}_{~\mu\rho}
-M^{\rho}_{~\alpha\rho}M^{\alpha}_{~\mu\nu}+M^{\rho}_{~\alpha\nu}M^{\alpha}_{~\mu\rho}~.
\ee
Finaly the curvature scalar of the physical spacetime 
\be 
R=g^{\mu\nu}\theta^{\rho}_a e^b_{\mu}\bar{R}^a_{~b\rho\nu}+D_{\mu}(M^{\rho\mu}_{~~~\rho}-M^{\mu\rho}_{~~~\rho})-M^{\rho}_{~\sigma\rho}M^{\sigma\mu}_{~~~\mu}+M^{\rho}_{~\sigma\mu}M^{\sigma\mu}_{~~~\rho}~,
\ee
should be 
\be
R=g^{\mu\nu}\theta^{\rho}_a e^b_{\mu}\bar{R}^a_{~b\rho\nu}+T_{\mu}T^{\mu}-\frac{1}{4}T_{\rho\sigma\mu}T^{\rho\sigma\mu}-\frac{1}{2}T_{\rho\sigma\mu}T^{\sigma\rho\mu}+2\nabla_{\mu}T^{\mu}=g^{\mu\nu}\theta^{\rho}_a e^b_{\mu}\bar{R}^a_{~b\rho\nu}+\mathbb{T}+2\nabla_{\mu}T^{\mu}~,
\ee
where $\mathbb{T}$ is precisely the one in MTG model, as defined in Eq. (\ref{tele1}). 
If the gravity theory for the physical spacetime is general relativity, the Einstein-Hilbert action $S=(1/2)\int d^4x\sqrt{-g}R$ after integrating out the divergence terms is 
\be\label{TGRlike}
S=\frac{1}{2}\int d^4x\sqrt{-g}(g^{\mu\nu}\theta^{\rho}_a e^b_{\mu}\bar{R}^a_{~b\rho\nu}+ \mathbb{T})~.
\ee
Now we have got a TGR like action for the spacetime perturbations. If the background spacetime is flat, $\bar{g}_{\mu\nu}=\eta_{\mu\nu}$,  $\bar{\omega}^a_{~b\mu}=0$, $\bar{R}^a_{~b\rho\nu}=0$ and $T^a_{~\mu\nu}=\partial_{\mu}e^a_{\nu}-\partial_{\nu}e^a_{\mu}$, the full action becomes $S=(1/2)\int d^4x\sqrt{-g}R$, going back to the action of TGR model (\ref{tgr}) under the Weitzenb\"{o}ck condition. 

Similar as before, in the action integral (\ref{TGRlike}), spacetime perturbation is not totally described by the second term $\sqrt{-g}\mathbb{T}$, the first term also contributes to the action for perturbation. We will again consider the expansion of the action as the perturbative series. 
To be consistent with the exponential map $g_{\mu\nu}=\left(e^{\epsilon}\right)^{\rho}_{~\mu}\left(e^{\epsilon}\right)^{\sigma}_{~\nu}\bar{g}_{\rho\sigma}$ introduced in the previous section, one should take the following map between the tetrads,
\be
e^a_{\mu}=\left(e^{\epsilon}\right)^{\rho}_{~\mu}\bar{e}^a_{\rho}~,~\theta^{\mu}_a=\left(e^{-\epsilon}\right)^{\mu}_{~\rho}\bar{\theta}^{\rho}_a~.
\ee
With these considerations one can obtain 
\be
\sqrt{-g}g^{\mu\nu}\theta^{\rho}_a e^b_{\mu}\bar{R}^a_{~b\rho\nu}=\sqrt{-\bar{g}}e^{\rm Tr\epsilon} \left(e^{-\epsilon}\right)^{\mu}_{~\rho}\left(e^{-\epsilon}\right)^{\nu}_{~\sigma}\bar{R}^{\rho\sigma}_{~~~\mu\nu}~.
\ee
Again, we know that $\mathbb{T}$ is in quadratic form of the torsion tensor $T^{\rho}_{~\mu\nu}$ and the latter is at least a first order perturbation quantity, so $\mathbb{T}$ should be at least a second order quantity. After expanding the action (\ref{TGRlike}) up to the second order: $S=S^{(0)}+S^{(1)}+S^{(2)}+\dots$, we find again that
$S^{(0)}=\frac{1}{2}\int d^4 x\sqrt{-\bar{g}}\bar{R}~,~S^{(1)}=-\int d^4 x\sqrt{-\bar{g}}\bar{G}^{\mu}_{~\nu} \epsilon^{\nu}_{~\mu}$. The former is the action for the background spacetime and leads to the background Einstein equation, the latter is the first order action which vanishes if the background equation is valid. The quadratic action from (\ref{TGRlike}) after using the background equation becomes
\be
S^{(2)}=\frac{1}{2}\int d^4 x\sqrt{-\bar{g}}[(\bar{R}^{\rho\sigma}_{~~~\mu\nu}+\bar{R}^{\rho}_{~\nu}\delta^{\sigma}_{~\mu}-\bar{R}^{\sigma}_{~\nu}\delta^{\rho}_{~\mu})\epsilon^{\mu}_{~\rho}\epsilon^{\nu}_{~\sigma}+\mathbb{T}]~.
\ee
This quadratic action is the same as Eq. (\ref{action2}) but in a different form. As before, all the derivatives of perturbations are contained in the torsion scalar $\mathbb{T}$, The background curvature, appearing as coefficients, merely contributes to the ``potential" of $\epsilon^{\mu}_{~\nu}$. The dynamics of the perturbation is mainly governed by torsion. 

\section{Conclusions}

Gravity is identical to curved spacetime and can be manifested by curvature, torsion or non-metricity. Armed with these multiple options, we in this paper revisited the problem of separating the physical spacetime into background and perturbation in perturbation theory, and considered the possibilities of formulating the gravitation of background and that of perturbation in separated ways. We showed that the perturbation to the curvature of a Riemannian spacetime can be represented in terms of non-metricity (in the metric formulation) or torsion (in the tetrad formulation), but the background is still of Riemannian geometry. With these separate treatments, we got  teleparallel like actions for the spacetime perturbation around a Riemannian background. 

Torsion and non-metricity have applications in other branches of physics. In the solid physics, topological defects caused by plastic deformations to the ideal crystals can be also formulated in the differential geometry. In this language, a kind of linear defects, called dislocations, is described by torsion (see the section 3.9 of Ref. \cite{Bahamonde:2021gfp}), and the point-like defects are described by non-metricity (see Ref. \cite{pointlike}  for an example). As an analogy, in this paper we want to provide a preliminary image in which the spacetime perturbations are considered as point-like topological defects or dislocations randomly distributed over the background spacetime. 

This formalism can be extended to some more general theories, such as the scalar-tensor theories in which the Lagrangian density is a linear function of the Ricci scalar, or the $f(R)$ theory which is equivalent to the former case after Legendre transformation and field redefinition. These theories were originally formulated in the so-called Jordan frame. However, through conformal transformation (or Weyl rescaling) they can be transformed into the Einstein frame in which the action of gravity has the form of Einstein-Hilbert, taking the price of introducing non-minimal couplings to the matter sector. Then the formulation presented here can be
applied to these theories in strait forward ways. 

{\it Acknowledgement}: This work is supported by the National Key R\&D Program of China Grant No. 2021YFC2203102 and by NSFC under Grant No. 12075231 and 12247103. 

{}

\end{document}